\documentclass[12pt]{iopart}
\usepackage{iopams}  
\usepackage{graphicx} 
\usepackage{epsfig}
\begin{document}

\title[]{Universal behavior of
baryons and mesons' transverse momentum distributions in the framework of percolation of strings.}

\author{L.Cunqueiro, J.Dias de Deus, E.G.Ferreiro and \underline{C.Pajares}$^1$}

\address{$^1$ IGFAE y Departamento de F\'{\i}sica de Part\'{\i}culas,
  Universidade de Santiago de Compostela, 15782-Santiago de Compostela,Spain}

\vspace{0.5cm}

The clustering of color sources\cite{Armesto:1996kt} reduces the average multiplicity and
enhances the average $<p_{T}>$ of an event in a factor $F(\eta)$ with respect
to those resulting from pure superposition of strings :
  \begin{equation}<\mu>=N_{s}F(\eta) <\mu>_{1}, \;
  <p_{T}^{2}>=<p_{T}^{2}>_{1}/{F(\eta)} \label{average} \end{equation}
where $N_{S}$ is the number of strings and $F(\eta)=\sqrt{\frac{1-e^{-\eta}}{\eta}}$ is a function of the density
of strings $\eta$\cite{Braun:2000hd}.
The invariant cross section  can be written as a superposition of the
transverse momentum distributions of each cluster, $f(x,p_{T})$ (Schwinger
formula for the decay of a cluster), weighted with
the distribution of the different tension of the clusters, $W(x)$ ($W(x)$ is
the gamma function whose width is proportional to $1/k$ where $k$ is a
determined  function of $\eta$ related to the measured dynamical transverse
momentum and multiplicity fluctuations)  \cite{DiasdeDeus:2003ei,Cunqueiro:2005hx}:
\begin{equation}\hspace{-1cm}\frac{dN}{dp_{T}^2 dy}=\int_{0}^{\infty}dx W(x)
f(p_{T},x)=\frac{dN}{dy}\frac{k-1}{k}\frac{1}{<p_{T}^2>_{1i}}F(\eta)\frac{1}{(1+\frac{F(\eta)p_{T}^{2}}{k<p_{T}^2>_{1i}})^{k}}.
 \label{spectra} \end{equation}

For (anti)baryons equation (\ref{average}) must be changed to 
$<\mu_{\overline{B}}>=N_{S}^{1+\alpha}F(\eta_{\overline{B}})<\mu_{1\overline{B}}>$
to take into account that baryons are enhanced over mesons in the fragmentation of a high density cluster. The parameter $\alpha=$0.09 is fixed from the
experimental dependence of $\frac{\overline{p}}{\pi}$ on $N_{part}$.
The (anti)baryons probe  higher densities than mesons,
$\eta_{B}=N_{S}^{\alpha}\eta$.  On the other hand, from the constituent
counting rules applied to the high $p_{T}$ behavior we deduce that for
baryons $k_{B}=k(\eta_{B})+1$.
In fig 1., we show the ratios $R_{CP}$ and $\frac{\overline{p}}{\pi^{0}}$
defined as usual, compared to RHIC experimental data for pions and antiprotons together with the LHC predictions. 
In fig.2 left we show  the nuclear modification factor $R_{AA}$ for
pions  and protons  for central collisions at RHIC. LHC predictions are also shown.
We note that pp collisions at LHC energies will reach enough string density
for 
nuclear like effects to occur. In this respect, in fig.2 , right, we show the ratio $R_{CP}$ for
$pp\to \pi X $ as a function of $p_{T}$, where the denominator is given by the
minimum bias inclusive cross section and the numerator is the inclusive cross
section corresponding to events with twice multiplicity than minimum bias.
According to our formula (\ref{spectra}) a suppression at large $p_{T}$
occurs.

We thank Ministerio de Educaci\'on y Ciencia of Spain under project FPA2005-01963and Conselleria de Educaci\'on da Xunta de Galicia for financial support. 

\begin{figure}
\begin{minipage}[t]{6cm}
\epsfig{figure=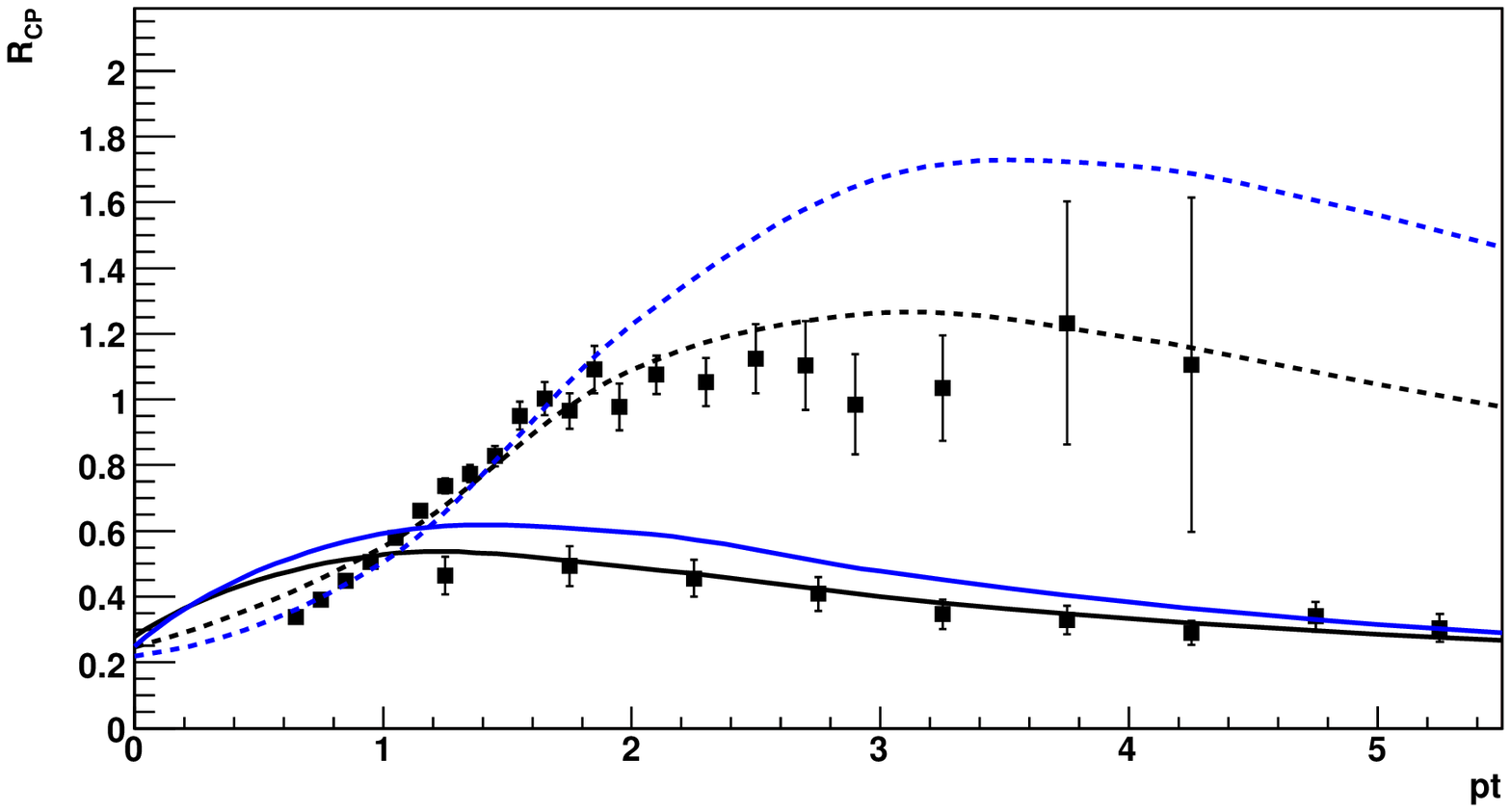,width=7cm,height=6cm}
\end{minipage} 
\hfill
\begin{minipage}[t]{6cm}
\hspace{-2cm} \epsfig{figure=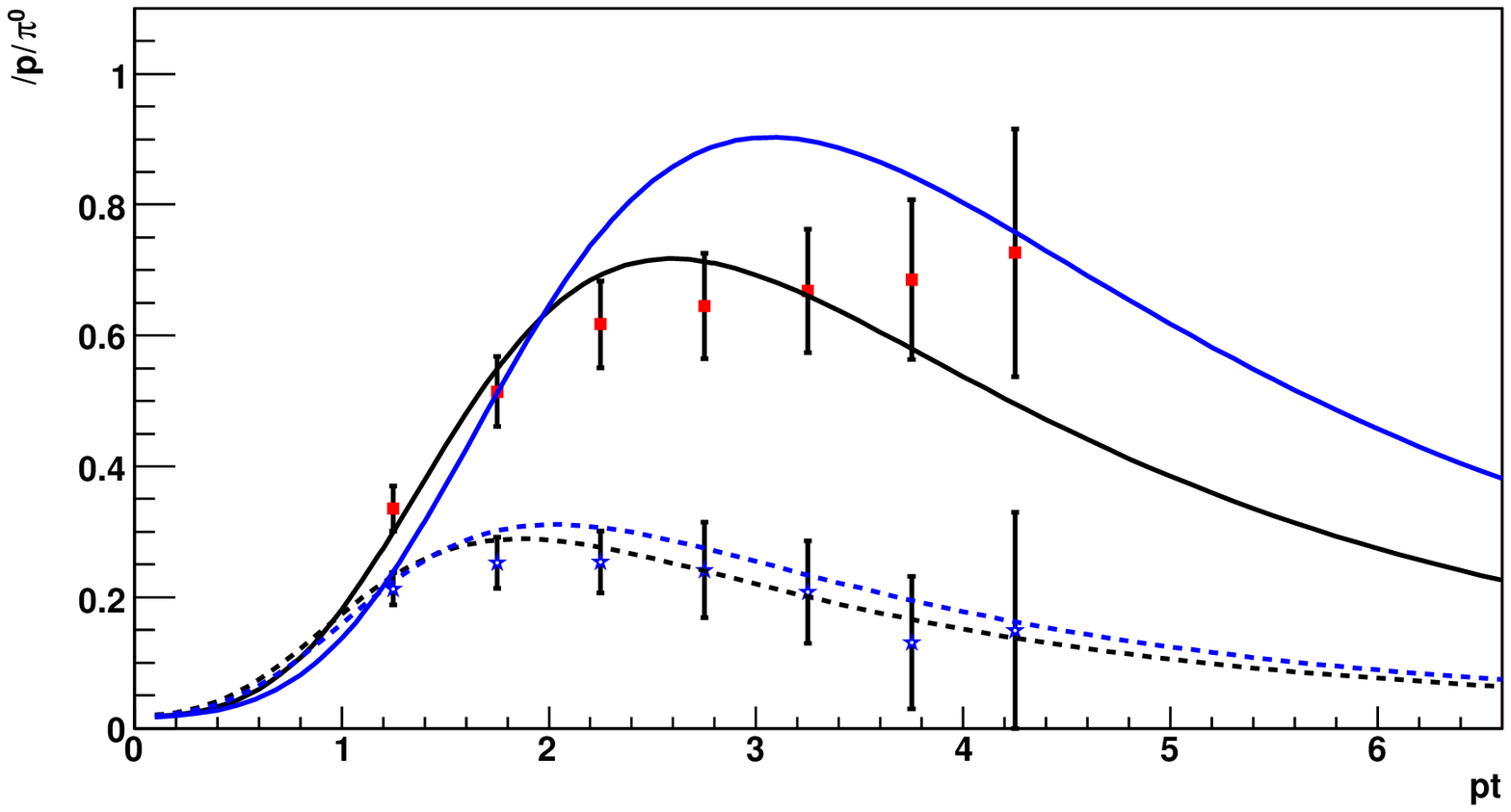,width=7cm,height=6cm}
\end{minipage} 
\caption{Left: $R_{CP}$ for neutral pions (solid) and
  antiprotons (dashed). Right: $\overline{p}$ to $\pi^{0}$ ratio for the centrality
  bins 0-10\% (solid) and 60-92\% (dashed). RHIC results in black and LHC predictions in blue. }
\end{figure}
\begin{figure}
\begin{minipage}[t]{6cm}
\epsfig{figure=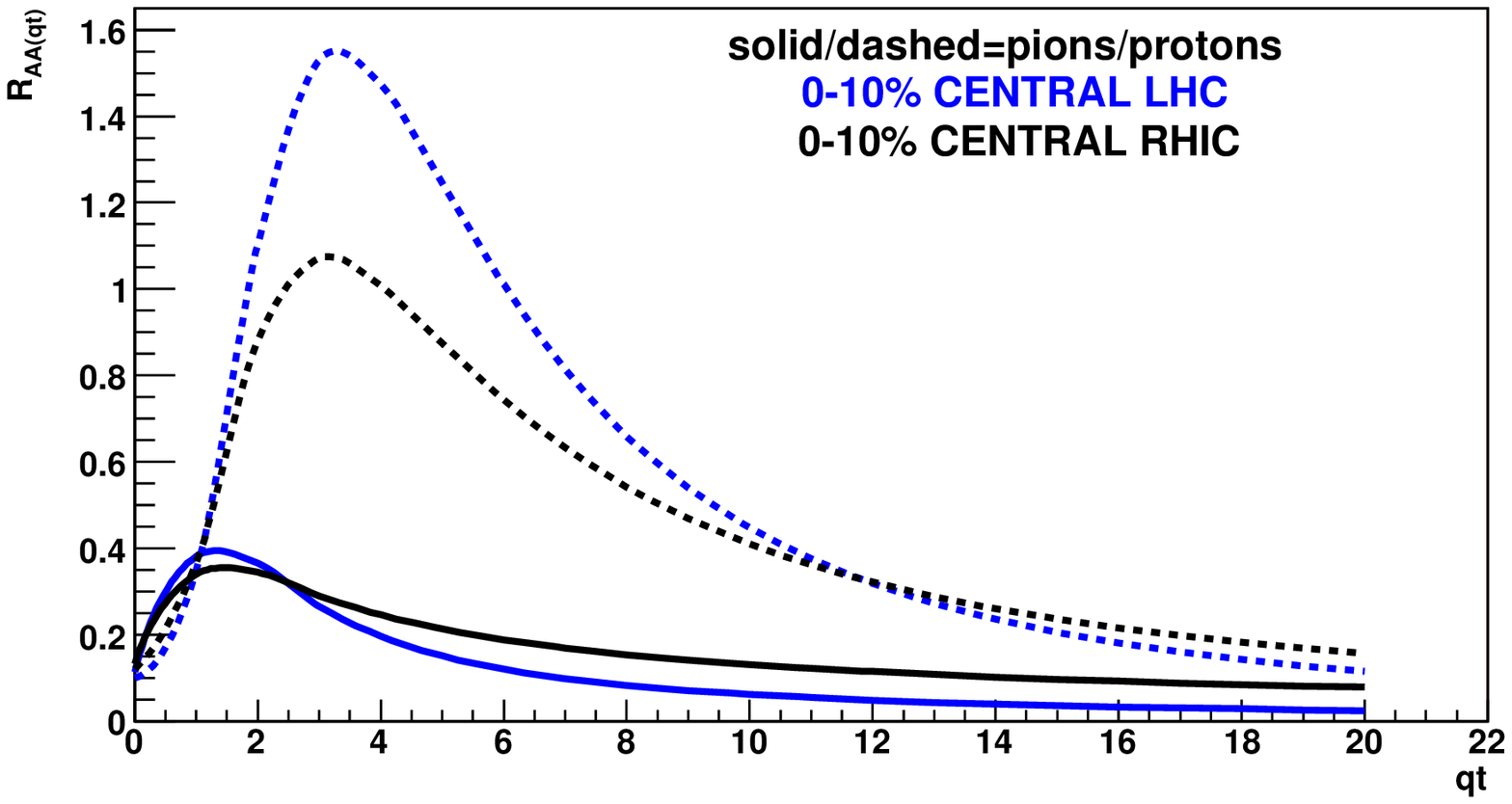,width=7cm,height=6cm}
\end{minipage} 
\hfill
\begin{minipage}[t]{6cm}
\hspace{-2cm} \epsfig{figure=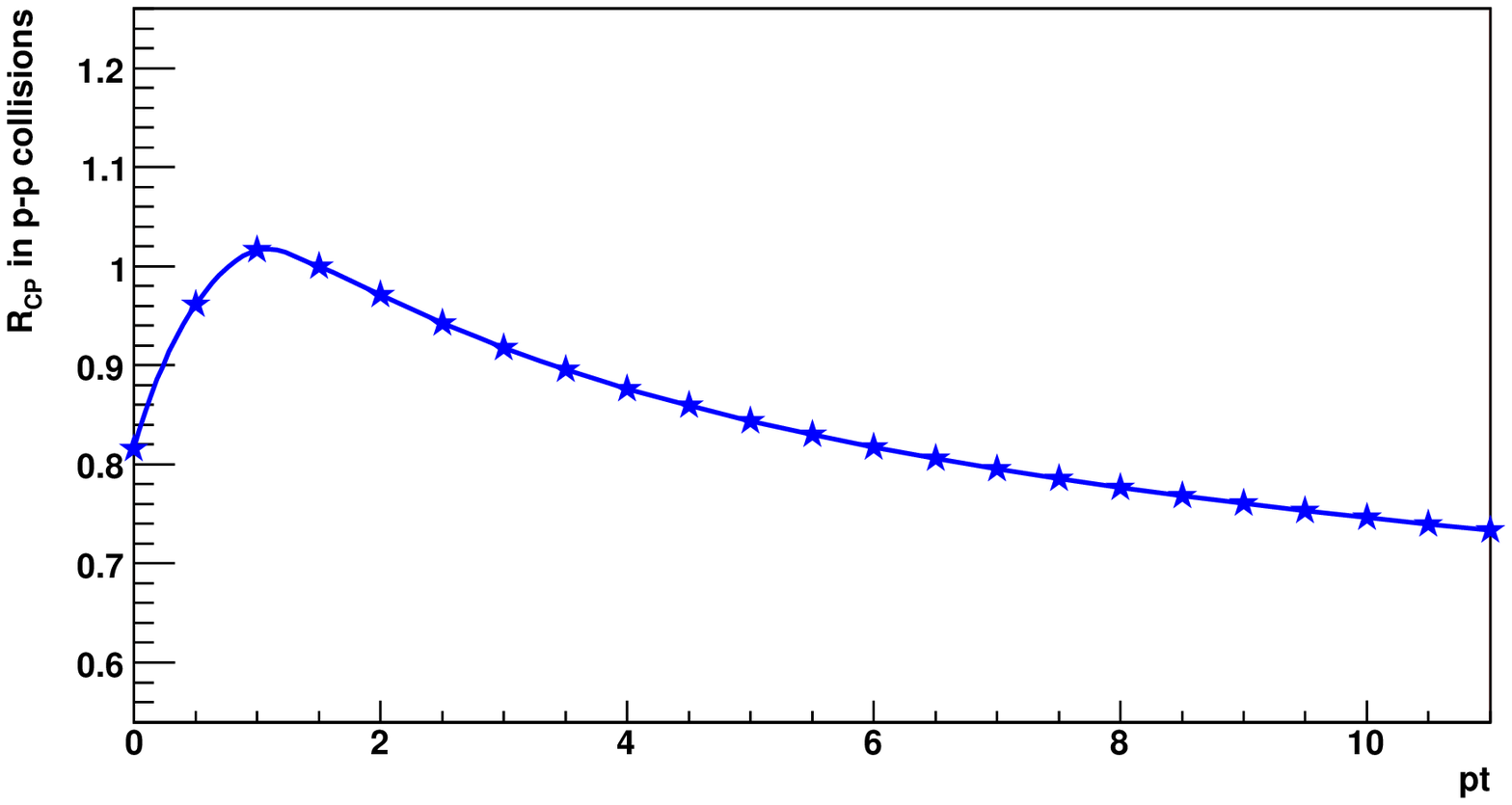,width=7cm,height=6cm}
\end{minipage} 
\caption{Left: Nuclear Modification Factor for $\pi^{0}$ (solid) and
  $\overline{p}$ (dashed) for 0-10\% central events, RHIC results in
  black and LHC predictions in blue. Right: $R_{CP}$ for pions in p-p collisions at LHC. }
\end{figure}


\section*{References}
\begin{thebibliography}{4}


\bibitem{Armesto:1996kt}
  N.~Armesto, M.~A.~Braun, E.~G.~Ferreiro and C.~Pajares,
  Phys.\ Rev.\ Lett.\  {\bf 77} (1996) 3736.

\bibitem{Braun:2000hd}
  M.~A.~Braun and C.~Pajares,
  Phys.\ Rev.\ Lett.\  {\bf 85} (2000) 4864.

\bibitem{DiasdeDeus:2003ei}
  J.~Dias de Deus, E.~G.~Ferreiro, C.~Pajares and R.~Ugoccioni,
  Eur.\ Phys.\ J.\  C {\bf 40} (2005) 229;  C.~Pajares,
  Eur.\ Phys.\ J.\  C {\bf 43} (2005) 9.

\bibitem{Cunqueiro:2005hx}
  E.~G.~Ferreiro, F.~del Moral and C.~Pajares,
  Phys.\ Rev.\  C {\bf 69}, 034901 (2004); \\ L.~Cunqueiro, E.~G.~Ferreiro, F.~del Moral and C.~Pajares,
  Phys.\ Rev.\  C {\bf 72} (2005) 024907.

\end{thebibliography}
\end{document}